\def\tabgamma{1}

\def\tabtaud{4}
\def\tabtauf{5}

\font\smallfont=cmr8

\font\Titlefont=cmssbx10 at 26pt
\font\Authorfont=cmssbx10 at 16pt
\font\Chapterfont=cmssbx10 at 20pt

\documentstyle[titlepage,12pt]{article}
\begin{document}
\begin{titlepage}
\begin{flushright}
HD-THEP-97-2\\
\end{flushright}
\vspace{1.5cm}

\begin{center}
{\Titlefont CP Violation}
\bigskip

{\Titlefont in radiative Z Decays}

\vspace{2cm}
\centerline{ \Authorfont D. Bru\ss,}

\vspace{0.3cm}
\centerline{Physics Department, University of Oxford}
\centerline{Clarendon Laboratory}
\centerline{Parks Road}
\centerline{Oxford OX1 3PU, UK}
\vspace{1.4cm}

\centerline{\Authorfont O. Nachtmann, P. Overmann}

\vspace{0.3cm}
\centerline{Institut f\"ur  Theoretische Physik}
\centerline{Universit\"at Heidelberg}
\centerline{Philosophenweg 16}
\centerline{D-69120 Heidelberg, FRG}
\vspace{2cm}

{\bf Abstract}\\
\parbox[t]{12cm}{ We propose to test the CP symmetry in the reactions
${\rm Z}\to\mu^+\mu^-\gamma$ and ${\rm Z}\to\tau^+\tau^-\gamma$.
The experimental analysis
of angular correlations allows to determine
a set of effective couplings:  the electric and weak dipole
moments of the muon and the tau lepton, and in particular chirality
conserving 4-particle couplings, all of which can be
induced by  CP--violation in renormalizable theories of
electroweak interactions beyond the
Standard Model. We update an indirect bound on the weak dipole moment of the
muon. }

\end{center}
\bigskip

\vbox{\hsize=10cm
\smallfont\baselineskip=10pt\noindent Research supported by
BMBF, contracts No. 05 6HD 93P(6), 05 7HD 91P(0), and by DFG, contract
No. Na 296/1-1. }
\end{titlepage}

\newpage

\bibliographystyle{unsrt}
\bigskip
\section{\Chapterfont Introduction}



\def\phat{\hat {\bf p}_+}
\def\khat{\hat{\bf k} }
\def\k{{\bf k}}
\def\q{{\bf q}}
\def\d{{\rm d}}
\def\Z{{\rm Z}}
\def\white{\vrule width 0pt height 16pt depth 8pt} 


The search for CP violation in Z boson decays provides an interesting tool
to investigate possible new physics
beyond the Standard Model. Several proposals
have been made how to look for CP violation in  processes such as
${\rm Z} \to \ell^+\ell^-,$ $\ell^+\ell^-\gamma$,
$\rm q \bar q,$ $\rm q \bar q \gamma$,
$\rm q\bar q G$, where $\ell = {\rm e},\mu,\tau$ leptons,
 q = u,d,s,c,b quarks,
and G = gluon (cf.\cite{Donoghue}--\cite{SomeRemarks} and
references therein).
CP--violating effects in the above reactions can be
parametrized by form factors, among them the
well known electric and weak
dipole moments of quarks and leptons, or one can use the effective
 Lagrangian approach.
Experimental investigations of CP--odd effects in Z decays
have been performed resulting in upper limits on
the weak dipole moment of the $\tau$ lepton \cite{Acton}--\cite{Stahl} and
on a CP violation parameter in $\rm Z \to b\bar b G$ \cite{ALEPHbbG}.
An ``indirect" limit on the $\tau$  electric dipole moment was given in  
\cite{Venturi}.

\medskip

Today, a large number of Z bosons
has been produced in electron--positron collisions
both at LEP (CERN) and at SLC (SLAC).
Thus investigations of relatively rare events like
radiative
three--body decays are possible.
In this article we investigate in detail the unpolarized
electron--positron annihilation to a   muon or tau lepton pair and a
hard photon with respect to CP violation:

\begin{equation}
{\rm e}^+(p_+)+ \enskip  {\rm e}^-(p_-) \enskip \to \enskip {\rm Z} \enskip
\to \enskip \ell^+(k_{\bar\ell})+ \enskip \ell^-(k_\ell)+
\enskip \gamma(q),\quad\quad(\ell = \mu, \tau).
\label{Reaction}
\end{equation}

Within the framework of the Standard Model (SM), CP violation  in these leptonic
reactions  occurs only at higher orders in perturbation theory, leading
to unobservably small effects.
However, interactions beyond the SM such
as the exchange of virtual photinos or other neutralinos,  
neutral Higgs bosons,  excited leptons, or
leptoquarks, can induce CP--violating contributions to  the
interaction of leptons and gauge bosons. A calculation of the dipole form factors
of the $\tau$ lepton in several extensions of the SM can be found in  
\cite{BeBraOv}.
Most of these models have in common that they lead to 
CP--violating effective couplings,
typically generated at  one--loop level, which are  
are not flavour--universal but can grow
substantially with the mass of the  quarks and leptons involved.

\medskip

In the following section a model--independent parametrization of CP--odd
couplings relevant for the reactions (\ref{Reaction}) is discussed. Then we propose
some easily measurable CP--odd momentum correlations which are optimized
to extract information on
these couplings from the data. In sections 4 and 5 we discuss in some detail
the level of accuracy of
CP tests obtainable
in Z decays to $\mu^+\mu^-\gamma$ and $\tau^+\tau^-\gamma$, respectively.

\newpage

\section{\Chapterfont Effective Lagrangian approach}

In order to describe in a model--independent way
possible CP--violating effects   we use an effective Lagrangian
with CP--odd operators of mass
dimensions $d > 4$.  Restricting ourselves to  $ d \le 6$
({\it after} symmetry breaking!),  the
operators which are relevant for the reactions considered here are given
 by  (c.f. \cite{BLMN,BN})

\begin{eqnarray}\label{LCP}
{\cal L}_{\rm CP}(x) && =\sum_\ell\left\{\right. \nonumber \\
-&&{i\over 2} \tilde d^\gamma_\ell \enskip
\bar\ell(x)\sigma^{\mu\nu} \gamma_5 \ell(x)
\left[ \partial_\mu A_\nu(x) - \partial_\nu A_\mu(x) \right] \hfill \\
-&&{i\over 2}  \tilde d^{\rm Z}_\ell \enskip
\bar\ell(x)\sigma^{\mu\nu} \gamma_5 \ell(x)
\left[ \partial_\mu Z_\nu(x) - \partial_\nu Z_\mu(x) \right] \nonumber \\
+&&\left[ f_{\rm V \ell} \bar\ell(x) \gamma^\nu \ell(x) +
f_{\rm A \ell} \bar\ell(x) \gamma^\nu  \gamma_5 \ell(x) \right]
Z^\mu(x) \left[ \partial_\mu A_\nu(x) - \partial_\nu A_\mu(x) \right]
\},
 \nonumber
\end{eqnarray}

\noindent
where $\tilde d_\ell^\gamma, \tilde d_\ell^Z$ are the electric
dipole moment (EDM) and weak dipole moment (WDM) coupling parameters
of the leptons $\ell = \mu, \tau$ and $f_{V\ell}, f_{A\ell}$ are real coupling
constants corresponding to CP--violating Z--photon--lepton
4-particle vertices. The photon and Z--boson fields are denoted by $A_\mu$ and
$Z_\mu$, respectively.
The authors of Ref. \cite{Lampe} have given a form factor decomposition
for reactions of the type (\ref{Reaction}).  In such a framework our
couplings in (\ref{LCP}) correspond in essence to the CP--violating
form factors with minimal momentum dependence.

\medskip

Following previous analyses \cite{BLMN,BBBHN} we introduce the
dimensionless parameters  $\hat f_{V\ell}, \hat f_{A\ell},
\hat d^\gamma_\ell, \hat d^{\rm Z}_\ell$ defined by

\begin{eqnarray}\label{Dimless}
f_{{\rm V}\ell, {\rm A}\ell} \enskip&=&\enskip -
\enskip{ e^2\cdot Q_\ell \over
\sin \theta_W \cos \theta_W m_{\rm Z} ^2 } \enskip
\hat f_{{\rm V}\ell,{\rm A}\ell} \hskip 3pt=
{2.62\cdot 10^{-5}\over \rm GeV^2}
\hat f_{{\rm V}\ell,{\rm A}\ell},\\
\tilde d_\ell^{\gamma,\rm Z} \enskip&=&\enskip -
\enskip{ e\cdot Q_\ell  \over
\sin \theta_W \cos \theta_W m_{\rm Z} } \enskip \hat d_\ell^{\gamma,\rm Z}
\quad =
{5.15 \cdot 10^{-16} e \rm cm} \cdot
\hat d_\ell^{\gamma,\rm Z},
\end{eqnarray}

\noindent where $\ell = \mu, \tau$ and $Q_\ell = -1$,
and $e^2/(4\pi)$ is the fine structure constant. 
 We use
$m_{\rm Z}=91.19$ GeV  and $\sin^2\theta_{\rm W}=0.23$ 
 for the numerics.
We  work to leading order in the CP--violating couplings of
${\cal L}_{CP}$  and in the couplings of the SM. Then the parameters in
the effective Lagrangian are identical to the corresponding form
factors, and imaginary parts of the form factors are absent.
This point was discussed in detail in \cite{SomeRemarks},
and it was shown that in this order one has

\begin{eqnarray}\label{5,6}
{\rm Re} \enskip d_\ell^{\gamma, \rm Z} (s) \quad &=&
\quad \tilde d_\ell^{\gamma, \rm Z}, \\
{\rm Im} \enskip d_\ell^{\gamma, \rm Z} (s) \quad &=& \quad 0.
\end{eqnarray}

\noindent Here $d_\ell^{\gamma, \rm Z} (s)$ are the EDM and WDM
{\it form factors} measurable in the reaction
$\rm e^+ e^- \to \ell^+ \ell^-$
at c.m. energy $\sqrt{s}$. We note that in a pure form factor approach
the form factors relevant for $\rm e^+ e^- \to \ell^+ \ell^-$ and
$\rm e^+ e^- \to \ell^+ \ell^- \gamma $ are {\it not} directly related.
Such a relation is only achieved after
making an assumption for the off--shell
dependence of the dipole moment form factors. On the other hand, in the
effective Lagrangian approach there is no problem to relate the parameters
of the above two reactions in leading order. Also in higher orders
it is in principle straightforward to work out this relation
(cf. \cite{SomeRemarks} for further remarks).

\medskip

The next step is to calculate the differential cross sections  for
$\rm e^+ e^- \to \ell^+\ell^-$ and
$\rm e^+ e^- \to \ell^+\ell^-\gamma$ taking into account the SM
diagrams (one is shown in Fig. 1a)
and the diagrams induced by the couplings of ${\cal L}_{\rm CP}$
(Figs. 1b,c,d).
Neglecting radiative decays with more than one $\gamma$ 
the sum of the two channels above gives the inclusive width

\begin{equation}\label{7}
\Gamma({\rm Z} \to \ell^+\ell^-X)
=
\Gamma_{\rm SM} ({\rm Z} \to \ell^+\ell^-X)
+ \Delta \Gamma_{\rm CP},
\end{equation}

\noindent
where $\Gamma_{\rm SM}$ is the SM width and $\Delta \Gamma_{\rm CP}$
is of second order
in the CP-violating
couplings of ${\cal L}_{\rm CP}$.
We use this calculation to make an {\it indirect}
estimate of the order of magnitude which is allowed for the
CP--violating parameters in ${\cal L}_{\rm CP}$ (cf. (\ref{LCP})).
The measured widths $\Gamma_{\mu\mu}$ and
$\Gamma_{\tau\tau}$
which include all radiative
events\footnote{We thank N. Wermes and M. Wunsch for
providing information concerning this point.}
 agree well with the predictions
from the SM (see Tab. 1).

\bigskip
\begin{center}
\begin{tabular}{|c|c|c|c|}\hline
\white \quad & $\Gamma_{\rm exp}$ &
 $\Gamma_{\rm SM}$  &$\sqrt{ \delta \Gamma_{\rm exp}^2 + 
\delta \Gamma_{\rm SM}^2 }$\\
 \hline \hline
\white $ \mu^+\mu^-X$    & 83.79 $\pm$ 0.22 MeV &
83.97 $\pm$ 0.01 $\pm$ 0.06 MeV& 0.23 MeV\\
\white $ \tau^+\tau^-X$  & 83.72 $\pm$ 0.26 MeV &
83.97 $\pm$ 0.01 $\pm$ 0.06 MeV&0.27 MeV\\
\hline\end{tabular}
\medskip

{{\bf Table \tabgamma}\quad Experimental and theoretical
(SM) values of the total
decay widths for $\rm Z \to \mu^+\mu^-X $ and $\rm Z \to \tau^+\tau^-X $
(cf. \cite{LEP,PDG}).}

\end{center}
\bigskip

\noindent We define values $\Delta\Gamma_{\rm exp}$
for the ``allowed" extra contribution to the width 
at $\lambda$ standard deviations as

\begin{equation}\label{8}
\Delta\Gamma_{\rm exp}^{\lambda \rm s.d.} \quad = \quad
\Gamma_{\rm exp} - \Gamma_{\rm SM} + \lambda \cdot
\sqrt{ \delta \Gamma_{\rm exp}^2 + \delta \Gamma_{\rm SM}^2 }.
\end{equation}

\noindent
In principle a combined estimate of the four CP--violation
parameters of (2) for each lepton $\ell$
is possible by ascribing to them any deviation from the width
as predicted by the SM, i.e. by setting $\Delta\Gamma_{\rm CP} \le  
\Delta\Gamma_{\rm exp}$.
However, this
would ignore the possibility of  other non--standard but CP--conserving  
effects to the widths
which can lead to an increase or decrease of the widths
as compared to the SM.
Thus such a procedure only gives an order of magnitude estimate of the
values allowed for the couplings in ${\cal L}_{\rm CP}$.
We give a simplified estimate by ignoring interference terms 
induced by the
new couplings, i.e. we assume
that only $f_{\rm V\ell}$ and $f_{\rm A\ell}$
or $\tilde d_\ell^{\gamma}$ or $\tilde d_\ell^{\rm Z}$ contribute in
the width $\Gamma({\rm Z} \to \ell^+\ell^-X)$.
As compared to the other couplings,
the contribution of the electric dipole moment
$\tilde d_\ell^{\gamma}$ is suppressed by a factor of $\alpha_{\rm ew}$
in the electroweak perturbation series.
Thus the leading terms in $\Delta\Gamma_{\rm CP},$
calculated without phase space cuts and neglecting the lepton masses,
are easily obtained from Tab. 1 of Ref. \cite{BBBHN} as follows

\begin{eqnarray}
\label{deltagammad}
\Delta\Gamma_{\rm CP}(\tilde d^{\rm Z}_\ell) &=&
\left ({|\tilde d_\ell^{\rm Z}| \over 10^{-17} \cdot e{\rm cm}}\right)^2
\cdot 0.24 {\rm\ MeV},\\
\label{deltagammaf}
\Delta\Gamma_{\rm CP}(\hat f_{\rm V\ell}, \hat f_{\rm A\ell}) &=&
\left( \hat f_{{\rm V} \ell}^2 + \hat f_{{\rm A} \ell}^2\right ) \cdot
0.042 {\rm\ MeV}.
\end{eqnarray}

\noindent
Now we can derive $\lambda$ s.d. limits on $\tilde d^{\rm Z}_\ell$ and
$\hat f_{\rm V\ell}, \hat f_{\rm A\ell}$ by requiring the theoretical widths
(\ref{deltagammad},\ref{deltagammaf}) to be smaller or equal 
$\Delta\Gamma_{\rm exp}^{\lambda\rm s.d.}$  (\ref{8}):

\begin{eqnarray}
\label{10a}
\Delta\Gamma_{\rm CP}(\tilde d^{\rm Z}_\ell) &\le&
\Delta\Gamma_{\rm exp}^{\lambda \rm s.d.},\\
\label{10b}
\Delta\Gamma_{\rm CP}(\hat f_{\rm V\ell}, \hat f_{\rm A\ell}) &\le&
\Delta\Gamma_{\rm exp}^{\lambda \rm s.d.},
\end{eqnarray}

\noindent We see from Tab. 1 that the differences of the central values,
$\Gamma_{\rm exp} - \Gamma_{\rm SM}$ are
negative.
Correspondingly the 1 s.d. values $\Delta\Gamma_{\rm exp}^{1 s.d.}$
are very small. Limits on 
$\tilde d_\ell^{\rm Z}, \hat f_{\rm V\ell}$ and $\hat f_{\rm A\ell}$
obtained from (\ref{10a}, \ref{10b}) at 1 s.d. are then not
to be trusted. As explained above, we should consider
(\ref{10a}, \ref{10b}) not as strict inequalities but only as
indications for the order of magnitude which is possible for the 
CP--violating couplings. We thus prefer to quote the 2 s.d. limits below,
since then $\Delta\Gamma_{\rm exp}^{2 s.d.}$ is at least of the order
of the combined experimental and theoretical errors given in the last
column of Tab. 1.

In this way we get from (\ref{10a}) 
for the WDM of the muon  the 2 s.d. upper limit

\begin{equation}\label{wdmindirect}
|\hat  d^{\rm Z}_\mu| < 0.021 \quad {\rm or}\quad |\tilde d^{\rm Z}_\mu| <
1.1\cdot10^{-17} e{\rm cm},
\end{equation}

\noindent while $\hat f_{\rm V\ell}$ and $\hat f_{\rm A\ell}$ are restricted
from (\ref{10b}) at the 2 s.d. level by

\begin{eqnarray}\label{findirect}
\hat f_{{\rm V} \mu}^2 + \hat f_{{\rm A} \mu}^2 &<& (2.6)^2, \\
\hat f_{{\rm V} \tau}^2 + \hat f_{{\rm A} \tau}^2 &<& (2.6)^2.\nonumber
\end{eqnarray}

\noindent 
This type of reasoning to obtain limits for the dipole moments was 
frequently used
(cf. e.g. \cite{Koerner,BBBHN}).
The estimate for $\hat d_\mu^{\rm Z} $ in (\ref{wdmindirect}) updates the one
given in \cite{Koerner}.

For the WDM of the $\tau$ lepton
it is not necessary to rely on indirect limits since experimental limits
exist, obtained from direct CP violation searches.
We quote the latest result for $|{\rm Re} d_\tau^{\rm Z}|$
which is related to our $\tilde d_\tau^Z$ as explained above:
The combination of all LEP results \cite{Wermes} gives the $95\%$ 
c.l. upper limits

\begin{equation}\label{13a}
|\hat d^Z_\tau|< 0.007
\quad {\rm or} \quad 
|{\rm Re} d_\tau^{\rm Z}| \enskip < \enskip 3.6\cdot 10^{-18} e{\rm cm}.
\end{equation}

\medskip

Estimates of the EDMs
which contribute at tree level to radiative events only (see Fig. 1)
can be given in a similar way by exploiting the experimentally allowed
range between the SM prediction and the measured values of
$\Gamma(\rm Z\to\ell^+\ell^-\gamma)$.
An upper limit was given for the $\tau$ EDM  in \cite{GriMen}.
Using the Z decay data collected up to now
it will be possible to update that value.
The EDM of the muon is known to be smaller
than $10^{-18} e\rm cm$, or  $|\hat d^\gamma_\mu| < 0.0019$ \cite{PDG}.
(Again, we can identify the on-shell EDM with the corresponding coupling
parameter in (2) for our purposes.)
The momentum correlations proposed in the following section
are blind to the EDM, therefore our analysis of $\rm Z\to\mu^+\mu^-\gamma$
aims at
only three still unknown parameters $f_{\rm V \mu}$,
$f_{\rm A \mu}$ and $\tilde d^{\rm Z}_\mu$.

\medskip

The calculation of the differential cross section shows that
it is convenient to choose instead of
the parameters
$f_{\rm V\ell}$ and $f_{\rm A\ell}$ the linear combinations

\begin{eqnarray}\label{f1f2}
\hat f_{1\ell} &=& g_{\rm V\ell}\cdot \hat f_{\rm A\ell} -  g_{\rm A\ell}
\cdot\hat f_{\rm V\ell},\\
\hat f_{2\ell} &=& g_{\rm V\ell} \cdot\hat f_{\rm V\ell} -  g_{\rm A\ell}
 \cdot\hat f_{\rm A\ell},\nonumber
\end{eqnarray}

\noindent where $g_{\rm V\ell} = T_{3\ell} - 2Q_\ell\sin^2\theta_W$
and $g_{\rm A\ell} = T_{3\ell} = -1/2$
are the usual neutral current coupling
constants of the leptons $\ell = \mu, \tau$.
In the zero--mass limit $\hat f_{1\ell}$  induces an  angular
distribution of the $\ell^+\ell^-\gamma$ momenta
which is measurable with tensor observables,
while $\hat f_{2\ell}$ is related to vector observables. Thus, with
appropriate observables these couplings can easily be determined independently.

\section{\Chapterfont Observables}

In the following we consider the case of unpolarized  $\rm e^+ e^-$ beams
in the reactions
(\ref{Reaction}). Then the initial state is described in the c.m. system
by a CP--invariant density matrix (CP tests with longitudinally polarized beams have been discussed in
\cite{BeBraOv,Rind}).
Furthermore we assume that the experimental phase space cuts
and the momentum reconstruction methods
do not introduce a CP bias.
Then the mean values of CP--odd momentum correlations trace the couplings
defined in ${\cal L}_{\rm CP}$.

\medskip

\medskip
The task is now to construct a set of momentum correlations whose
distributions are sensitive to the couplings  defined in ${\cal L}_{\rm CP}$ with
a reasonable statistical significance.
For simplicity, we first consider each of the parameters $\hat{f_{1\ell}}$,  
$\hat{f_{2\ell}}$, $\hat d^{\rm Z}_\ell$ and
$\hat {d^\gamma_\ell}$
separately by setting the other three to zero.
Then the linear contribution to the mean value of a given observable,
together with the width of its distribution,
 yields an estimate of the statistical
accuracy with which this coupling can be measured.

\medskip

The kinematics of the reaction $\rm Z \to \mu^+\mu^-\gamma$
is determined by the $\mu^+, \mu^-$ momenta $\bf k_\pm$ in the
Z rest frame which is to a good approximation the
laboratory system at LEP and SLC.
In the case of $\rm Z \to \tau^+\tau^-\gamma$ with subsequent
one--prong $\tau$ decays which are the most frequent ones
(about 85 \%), the measurable quantities
are primarily the momenta of the charged  $\tau^\pm$ decay products, again  
denoted by
$\bf k_\pm$, and the photon momentum $\bf q$.
Using these momenta one can construct many CP-odd observables.
We have calculated the signal--to--noise ratios of some basic
correlations and found the following observables
especially useful:

\noindent

\begin{eqnarray} \label{Observables}
T &=& (\khat_+ - \khat_-)\cdot\phat \enskip
 {(\khat_+ \times \khat_-)\cdot\phat},\\
V &=&  {(\khat_+ \times \khat_-)\cdot\phat}.\nonumber
\end{eqnarray}

\noindent Here hats denote unit momenta, and $\bf p_+$
is the initial positron momentum vector. With respect to the final state
momenta $T$ is a component of a tensor and $V$ a component of a
 vector observable.
These correlations  involve neither particle energies nor the
photon momentum.
For $\rm Z \to \tau^+\tau^-\gamma$ we additionally  studied  the observables  
$T^*$ and
$V^*$ which are defined as above but with $\hat {\bf k}_\pm$ taken
in the  rest system of the $\tau^+\tau^-$ pair.
The $\tau$ momenta usually
cannot be fully reconstructed from the observed decay products and
the $\tau$ decay vertex information. But the observation of the photon
should be sufficient to reconstruct in each event the $\tau^+\tau^-$
center of mass frame.

\medskip
Let us denote the coupling parameters $\hat f_{1\ell},\hat f_{2\ell},
\hat d^Z_\ell,\hat d_\ell^\gamma$ (cf. (2-4, \ref{f1f2})) by
$g_{1\ell},..., g_{4\ell}$.
The mean values of CP--odd observables such as $T$ and $V$ defined above
are in general not purely linear functions of
the $g_{i\ell}$ since the
normalization introduces a denominator with
quadratic contributions in the $g_{i\ell}$'s.
The second order terms certainly are negligible
if the couplings are small. For practical purposes
it might be useful to consider observables which get
{\it linear contributions only}. These can easily be obtained by
multiplying any of the momentum correlations discussed in this article
with the measured width $\Gamma(Z\to\ell^+\ell^-\gamma)$
for the phase space region, i.e. the cuts considered.
In this way one constructs
observables whose mean values are strictly linear in the
couplings to be measured:

\begin{equation}
\Gamma({\rm Z\to\ell^+\ell^-\gamma})\enskip\cdot\enskip
 \langle {\cal O} \rangle \quad = \quad
\sum_{ i = 1}^4 a_i \cdot g_{i\ell}.
\end{equation}
The $a_i$ are proportionality constants which can
be calculated numerically, and $\cal O$ denotes any CP--odd observable
which does not depend on the unknown couplings.

\medskip

In order to get an impression about the quality of the observables
introduced above we also investigate for the couplings
$g_{i\ell}$ the corresponding
{\it optimal observables} which have the greatest possible statistical
signal--to--noise ratio. For a single coupling this method was introduced
in \cite{Atwood,Davier}.
The generalization to an arbitrary number of new couplings
was given in \cite{Diehl} (cf. also \cite{DN}).
The optimal observables are constructed as follows:
Let us write the differential cross section as

\begin{equation}\label{17}
\d\sigma(\phi) = \d\sigma_{\rm SM}(\phi) 
+ \sum_{i=1}^4 \d\sigma_i^{(1)}(\phi)
g_{i\ell}+\sum^4_{i,j=1}\d\sigma_{ij}^{(2)}(\phi)g_{i\ell}g_{j\ell}.
\end{equation}

\noindent Here $\phi$ stands collectively for the phase
space variables, $\d \sigma_{\rm SM}$
denotes the CP--conserving SM part of the differential cross section, and
$\d\sigma_i^{(1)}$ are the contributions due to the interference
of the CP--odd interactions of ${\cal L}_{\rm CP}$ with the SM amplitudes.
 Then the ratios

\begin{equation}\label{OO}
{\cal O}_i(\phi) =
 { \d\sigma_i^{(1)}(\phi) \over
\d\sigma_{\rm SM}(\phi) },
\quad\quad\quad (1\leq i\leq 4)
\end{equation}

\noindent are the observables with optimal statistical sensitivity for a joint
measurement of the couplings assuming that the couplings which shall be  
measured are small.
The observables  ${\cal O}_i$ (\ref{OO}) are optimal for arbitrary
 phase space cuts.  
Their sensitivity, however,  depends on the cuts.

For the reaction $\rm Z \to \mu^+\mu^-\gamma$ one can easily
measure the momenta of all final state particles.
Let $q$ be the $\gamma$ momentum and 
$k_{\bar \mu} \equiv k_+, k_\mu \equiv k_-$
the momenta of $\mu_\pm$.
Thus in this case the phase space variable $\phi$ in (\ref{OO}) 
is given by

\begin{equation}\label{muphasespace}
\phi \equiv ( \bf k_+, k_-, q).
\end{equation}

\noindent Due to the energy--momentum conservation constraint
the phase space is 5--dimensional. Identification of a photon 
requires some isolation cuts. 
For our analysis we
assume covariant cuts of the form

\begin{equation}
\label{ycut}
{ (k_{(1)} + k_{(2)})^2 \over m_{\rm Z}^2 } \ge y,
\end{equation}

\noindent where
$k_{(1)}, k_{(2)}$ are any two different
four momenta from the set $k_+, k_-, q$,
and $y$ is chosen to be $0.03$.

\medskip

For the reaction $\rm Z\to \tau^+\tau^-\gamma$ with subsequent
1--prong $\tau$ decays 

\begin{equation}
{\rm e}^+(p_+) + {\rm e}^-(p_-) \to {\rm Z} \to 
\tau^+(k_{\bar\tau}) + \tau^-(k_\tau) + \gamma(q),
\end{equation}

\begin{eqnarray} \label{channels}
\tau^+(k_{\bar\tau}) &\to& a^+(k_+) + {\rm neutrals},\\ \nonumber
\tau^-(k_\tau) &\to& b^-(k_-) + {\rm neutrals}, 
\quad (a,b = \pi,\rho,\rm e,\mu) 
\end{eqnarray}

\noindent we assume integration over the phase space variables of the
neutrals. We are then left ideally with the information
from the momenta of $\tau^+, \tau^-, \gamma, a^+, b^-$ and thus here we can 
identify

\begin{equation}\label{tauphasespace}
\phi \equiv (\bf k_{\bar\tau}, k_\tau, k_+, k_-, q ).
\end{equation}

\noindent Of course, these momenta are not totally independent.
Energy--momentum conservation in production and in the decays
imposes well known constraints. 
 Depending on the decay channels (\ref{channels}) the
phase space is 9 to 11 dimensional.
For our analysis below we assume
again the isolation cuts (\ref{ycut}) for the momenta of the
$\gamma$ and the charged decay products $a^+, b^-$. In practice
the complete information on the $\tau^\pm$ momenta may not always
be available. We will, therefore, discuss the optimal observables for
the ideal case where the $\tau^\pm$ momenta can be reconstructed.
This then gives the ideal sensitivities. On the other hand,
for the simple observables $T, V$ of (\ref{Observables})
and their analogues $T^*, V^*$ constructed in the
$\tau^+\tau^-$ c.m. system one does not need knowledge of the
individual $\tau^+$ and $\tau^-$ momenta. Obviously, with partial
information on the $\tau^\pm$ momenta one can then reach
sensitivities in between. 

\medskip

From (\ref{OO}) 
we get as many optimal observables as we have parameters.
Of course, in general the expectation values
$\langle {\cal O}_i \rangle$ will get linear
 contributions from all four parameters in leading order.
We have investigated the matrix describing this dependence of the mean values
of observables on the parameters. For the optimal observables this is 
in leading order identical 
to the correlation matrix of the ${\cal O}_i$ (cf.\cite{Diehl}). We found that
these correlations are either negligible, or the
couplings contributing  are bounded by independent experiments.
In the following we therefore confine ourselves to discuss the diagonal elements of 
the covariance matrix.
We calculated numerically the mean values and the widths
of the  ${\cal O}_i$ and got in this way for a given number of events the
statistical accuracy with which each of the  unknown parameters can be
measured ideally, assuming that all other new couplings are zero.
For a numerical evaluation it is not necessary
to have the ${\cal O}_i(\phi)$ in explicit form.
We use a Monte Carlo event generator\footnote{
Monte Carlo event generators for the reactions ${\rm e^+e^- }\to\mu^+\mu^-\gamma$  
and ${\rm e^+e^-}\to\tau^+\tau^-\gamma$ with  subsequent $\tau$ decays
are available from the authors.} for the reactions
(\ref{Reaction}) which takes into account the effects of ${\cal L}_{\rm CP}$.
Since the optimal observables are composed from terms of the differential
cross section, one can easily generate them numerically with a Monte Carlo
method.

\medskip

The knowledge of the optimal observables is interesting
by itself, and useful for deriving simpler correlations which are
approximately optimal. In  most cases we found such observables
which presumably are preferable for an experimental
analysis. 

\bigskip\bigskip

\section{\Chapterfont Z decays to $\mu^+ \mu^- \gamma$}

The LEP experiments have collected several thousand events of the reaction
${\rm e^+ e^-} \to \mu^+ \mu^- \gamma.$ For the CP tests proposed here
it is sufficient to measure the muon directions of flight ${\hat\k}_\pm$.
Using these one can calculate the mean values of the observables
$T$ and $V$ defined in (\ref{Observables}).
The tensor correlation $T$ gets a contribution only from
$\hat f_{1\mu},$ allowing a clean determination of this parameter.
On the other hand, the mean value of the vector observable $V$ depends
both on $\hat f_{2\mu}$ and $\tilde d_\mu^{\rm Z}.$  As expected, the
vector correlation is slightly less sensitive than the tensor observable, due
to the smallness of the vector polarization of the Z boson.
The contribution of the weak dipole moment to $V$ is negligible: in order
to cause a measurable effect it would have to be three orders of magnitude bigger 
than the indirect upper limit (\ref{wdmindirect}).
We conclude  that  $\hat f_{2\mu}$ can be measured practically without  
contamination
by other CP--violating  sources using the observable
$V$ of (\ref{Observables}).

\medskip

We also investigated the correlation of a measurement of our
parameters with optimal observables. Since the EDM coupling
$\hat d^\gamma_\mu$ does not contribute to CP-odd terms in
leading order (cf.\cite{BLMN}), we have here three parameters
$\hat f_{1\mu},\hat f_{2\mu}$, $\hat d_\mu^Z$ (denoted by
$g_{1\mu},..., g_{3\mu}$)
and three observables ${\cal O}_i$ (cf. (\ref{OO})). Working
always in leading order in the $g_{i\mu}$, we have

\begin{equation}\label{Oi}
\langle{\cal O}_i\rangle=\sum^3_{j=1}c_{ij} \cdot g_{j\mu}
\end{equation}

\noindent with

\begin{equation}
\label{cij}
c_{ij}=\langle{\cal O}_i{\cal O}_j\rangle_{\rm SM}.
\end{equation}

\noindent We calculated this covariance matrix. In Tab. 2 we give
the $c_{ii}$ ($i$=1,2,3) and the rescaled matrix

\begin{equation}\label{19c}
\langle{\cal O}_i{\cal O}_j\rangle_{\rm SM}/\sqrt{c_{ii}c_{jj}},
\end{equation}

\noindent where the diagonal elements are equal to 1.
From these numbers we can easily derive parameters
$g'_i$ $(i=1,2,3)$ where the corresponding optimal observables
${\cal O}'_i$
have a diagonal covariance matrix
$c'_{ij} = \langle {\cal O}'_i {\cal O}'_j\rangle_{\rm SM}$.
These parameters can be chosen as

\begin{eqnarray}
\nonumber
g'_1 &=& \hat f_{1\mu},\\
\label{23a}
g'_2 &=& \hat f_{2\mu} - 1.1\cdot 10^{-4} \hat d_\mu^{\rm Z}, \\
\nonumber
g'_3 &=& \hat d_\mu^{\rm Z}.
\end{eqnarray}

\noindent Recall that for optimal observables the covariance
matrix $V(g')$ of the estimated couplings is the inverse of 
the number of events times the covariance matrix of the
observables (cf. \cite{Diehl, DN})

\begin{equation}
\label{23b}
V(g') = (N c')^{-1}.
\end{equation}

\noindent Thus a diagonal $c'$ ensures a diagonal $V(g')$.
 We conclude from (\ref{23a}) that also for
optimal observables the correlations are
negligible. 
For a general discussion of diagonalization
procedures in the estimation of more than one
coupling parameter we refer to \cite{DN}.

\medskip

From the above we find that 
 we have to consider for any of our observables ${\cal O}=T,V$ or an
optimal one only a dependence on one coupling parameter. Calling
it generically $g$ we get in linear approximation

\begin{equation}\label{19d}
\langle{\cal O}\rangle=c \cdot g.\end{equation}

The 1 s.d. accuracy obtainable for a determination
of $g$ by a measurement of ${\cal O}$ can be estimated as

\begin{equation}\label{deltag}
\delta g=\frac{\sqrt{\langle{\cal O}^2\rangle_{\rm SM}}}
{|c|\sqrt{N}},\end{equation}

\noindent where $\langle {\cal O}^2\rangle_{\rm SM}$
is the variance of ${\cal O}$, calculated in the SM,
and $N$
is the number of events within the cuts considered.

\medskip

In Tab. 3 we summarize
the statistical accuracies $\delta g$ of (\ref{deltag})
with which the WDM of the muon,
 $\hat f_{1\mu}$ and $\hat f_{2\mu}$  can be measured assuming a
 sample of $N = 10^4$ $\mu^+\mu^-\gamma$ events within
the cut (\ref{ycut}). The numbers
in Tab. 3
 represent the minimal sizes of the couplings which lead to
visible effects at 1 s.d., or, if no deviation from the
CP symmetry is found, $\delta g$ is the 1 s.d. upper limit
on the coupling parameter.
The upper limits on $\hat f_{1\mu}, \hat f_{2\mu}$
correspond to two stripes in
the $\hat f_{\rm V\mu}, \hat f_{\rm A\mu}$ parameter space.
These are depicted in Fig. 2,
together with the indirect bound (\ref{findirect}) assuming
zero mean values and  a phase space cut parameter $y = 0.03$.
The  sensitivities 
of the optimal observables for $\hat f_{1\mu}$ and $\hat f_{2\mu}$
 are plotted as a function
 of the cut parameter $y$ in Fig. 3.

\bigskip

\begin{center}

\hbox{
\begin{tabular}{|c|c|c|}\hline
  $c_{11}$ & $c_{22}$ & $ c_{33} $ \white \\ \hline\hline
\white $2.64\cdot 10^{-3}$ & $5.61\cdot10^{-4}$ & $1.17\cdot 10^{-8}$\\
\hline\end{tabular}
\hskip 1cm \hfill
\begin{tabular}{|c||c|c|c|}\hline
 & $\hat f_{1\mu}$ & $\hat f_{2\mu}$ & $ \hat d_\mu^{\rm Z} $ \white \\ \hline\hline
\white $\hat f_{1\mu}$        &1&0.000&0.000\\\hline
\white $\hat f_{2\mu}$        &0.000&1&-0.023\\ \hline
\white $\hat d_\mu^{\rm Z}$   &0.000&-0.023&1\\
\hline\end{tabular}
}

\bigskip

\begin{center}
{{\bf Table 2}\quad The variances $c_{ii}=\langle{\cal O}_i
{\cal O}_i\rangle_{\rm SM}$ and the rescaled covariance
matrix
$\langle {\cal O}_i {\cal O}_j\rangle_{\rm SM}(c_{ii}
c_{jj})^{-1/2}$ of the optimal observables
for the measurement of the parameters
 $\hat f_{1,2\mu}$ and  $ \hat d_\mu^{\rm Z}$. }
\end{center}

\bigskip

\begin{tabular}{|c||c|c|c|}\hline
 & $\hat f_{1\mu}$ & $\hat f_{2\mu}$ & $ \hat d_\mu^{\rm Z} $ \white \\ \hline\hline
\white optimal observable   &\quad 0.195\quad&\quad 0.423\quad &\quad 92.5\quad\\
\hline
\white $T$    &\quad 0.238\quad  &&\\ \hline
\white $V$    &&\quad 0.523\quad &\quad 101\quad\\
\hline\end{tabular}

\bigskip

{{\bf Table 3}\quad 1 s.d. accuracies
$\delta g$ (\ref{deltag}) with which the couplings $g=\hat f_{1\mu}$,
$\hat f_{\rm 2\mu}$, and $\hat d^{\rm Z}_\mu$ can be measured  assuming a
sample of 10000 decays ${\rm Z} \to \mu^+\mu^-\gamma$
within the cut (\ref{ycut}). Missing entries
correspond to a vanishing sensitivity ($\delta g=\infty$).}
\end{center}

\bigskip

We note that a search for CP--odd effects in the two--body decay
 ${\rm Z} \to \mu^+ \mu^-$
is in practice not possible \cite{BLMN} since the direction of flight of the
muons  carries no
CP--odd information, and the spin analyzing muon decays usually
escape detection. Nevertheless, this reaction should be
useful to study possible detector effects which may feign
non--zero mean values or asymmetric distributions of the
correlations (\ref{Observables}).

\bigskip\bigskip

\section{\Chapterfont Z decays to $\tau^+ \tau^- \gamma$}

Compared to the muon case, CP tests in Z decays to
$\tau^+ \tau^- \gamma$ follow a somewhat different line:
The $\tau$ momenta can not (yet) be measured directly.
Instead we use the momenta of the charged decay products
of the $\tau$ leptons to build the
CP--odd correlations $T,V$ and their
counterparts $T^*, V^*$ which use the momenta
in the $\tau^+\tau^-$ rest frame.
As explained in Sect. 3 we assume, however, knowledge of 
the $\tau^\pm$ momenta for the optimal observables.

\medskip

In order to cover a large fraction (about 35\%) 
of the $\tau^+ \tau^-$
events, we investigate all combinations of the following
one--prong decay channels:

\begin{eqnarray}
\label{Pi}\tau^\pm &\to& \nu \pi^\pm(k_\pm), \\
\label{Rho} &\to& \nu\rho^\pm(k_\pm), \\
\label{Lepton}&\to& \nu \bar\nu \ell^\pm ( k_\pm ) \quad\quad (\ell =
{\rm e},\mu) .
\end{eqnarray}

\noindent For these decays the SM matrix elements are assumed.

\medskip

In Tables \tabtaud,\tabtauf\ we collect the upper
limits on the CP--odd  couplings
which can be obtained by measuring the
optimal observables and the correlations $T, V$,
$T^*,$ and $V^*$. 
The numbers $N$ of events for the various channels correspond
roughly to a total number of $10^4$ ${\rm Z} \to \tau^+\tau^-\gamma$
events within the cut (\ref{ycut}).
Here we always assume that only one parameter
$g=\hat d^\gamma_\tau,\hat d^Z_\tau,\hat f_{1\tau},\hat f
_{2\tau}$ is different from zero and quote $\delta g$ calculated
according to (\ref{19d}, \ref{deltag}) for the observable indicated.
The measurement of the dipole moments relies essentially
on the analysis of the $\tau$ spins.
The momentum distribution of the final state particles depends
on the $\tau$ polarization. For instance, the $\pi^+$ in
the reaction (\ref{Pi}) is  emitted preferentially opposite to the
$\tau^+$ spin direction, if we consider the decay in the $\tau^+$
rest system.
The momentum correlations
$T, V, T^*, V^*$
 which we use project on the CP--asymmetric
part of the spin-spin-correlation of the $\tau$ pair.
Therefore they are rather insensitive to the electric dipole coupling
$\tilde d^\gamma_\tau$
which affects the polarization of only one $\tau$ lepton
(see Fig. 1c). The weak dipole coupling $\tilde d_\tau^{\rm Z}$,
however, can be measured
with an accuracy of the order of $10^{-17} e$cm assuming
that $10^4$ $\rm Z \to \tau^+\tau^-\gamma$ decays are available.
The obvious gap between the sensitivities of the
optimal observables and the
simple correlations originates from the fact that the optimal observables
as we calculate them work with the knowledge of the $\tau$ momentum.
At least for the measurement of the electric dipole moment this seems
to be a basic ingredient. Since a measurement of the EDM and
WDM is based on the $\tau$-spin information, we observe a rather
strong dependence of the sensitivities in Tab. 4 on the $\tau$-decay
channels, corresponding to their spin-analyzing qualities.

\medskip

The  parameters $\hat f_{\rm 1\tau},$ $\hat f_{\rm 2\tau}$ can be determined
with an accuracy of 1 s.d. when they are of the order of about 0.5
(Tab. \tabtauf), given the assumed number of events.
Here it should be advantageous to use
the observables $T$ and $V$ which get only
tiny contributions from the dipole moments of the tau lepton.
These results are comparable to the muon case (Tab. 3).  
It is worth noting
that the sensitivities are more or less independent of the tau decay channel:
the measurement of $\hat f_{1\tau}, \hat f_{2\tau}$  
is not substantially based on the spin analyzing 
qualities of the different  decay modes.

\bigskip

\font\small=cmr9

\begin{center}
\small

\begin{tabular}{|c|c||c|c|c|c|c||c|c|c|c|c|}\hline
 \multicolumn{2}{|c||}{}&
 \multicolumn{5}{|c||}{$\hat d_\tau^\gamma$}&
 \multicolumn{5}{|c|}{$\hat d_\tau^Z$}  \white \\ \hline
\white
channel & $N$ & opt. & $T$ & $V$ & $T^*$ & $V^*$ &
 opt. &$T$ & $V$ & $T^*$ & $V^*$ \\ \hline \hline
\white$\pi-\pi$   & 100 &0.40&91&59&8.4&5.0&0.054&0.83&8.3&0.077&0.67\\
\white$\pi-\rho$  & 400 &0.41&67&41&13&3.2&0.038&0.63&38&0.051&1.8\\
\white$\rho-\rho$ & 400 &0.85&150&83&42&5.9 &0.070&1.3&28&0.092&44\\
\white$\ell-\ell$ & 800 &0.41&74&51&14&4.8 &0.044&0.73&12&0.063&0.63\\
\white$\ell-\pi$  & 600 &0.26&120&95&7.5&17 &0.030&0.98&3.4&0.140&0.46\\
\white$\ell-\rho$ &1200 &1.5&1400&810&53&40&0.13&38&23&0.91&4.3\\
\hline \hline
\multicolumn{2}{|c||}{combined}\white&\bf 0.17&40&26&4.8&\bf 2.2 &
\bf 0.018&0.37&3.0&\bf 0.032&0.32\\
\hline
\end{tabular}

\bigskip

\noindent{\bf Table \tabtaud}{\rm\quad 1 s.d. accuracies with which  $\hat  
d_\tau^\gamma$ and
$\hat d_\tau^{\rm Z}$ can be measured \\ for a given number $N$
of events in various
$\tau$ decay modes}

\bigskip
\begin{tabular}{|c|c||c|c|c|c|c||c|c|c|c|c|}\hline
 \multicolumn{2}{|c||}{}&
 \multicolumn{5}{|c||}{$\hat f_{1\tau}$}&
 \multicolumn{5}{|c|}{$\hat f_{2\tau}$}  \white \\ \hline
\white
channel & $N$ & opt. & $T$ & $V$ & $T^*$ & $V^*$ &
 opt. &$T$ & $V$ & $T^*$ & $V^*$ \\ \hline \hline
\white$\pi-\pi$   & 100   &0.61&1.7&7.1&104 &45&1.1&3.4&3.7&36&31\\
\white$\pi-\rho$  &400  &0.53&0.85&5.9&59&32&0.88&2.9&1.9&38&36\\
\white$\rho-\rho$ & 400 & 0.70&0.88&17&60&57&1.3&8.6&2.0&46&68\\
\white$\ell-\ell$ & 800   &0.48 &0.60&4.8&29&62&0.87&2.2&1.4&22&34\\
\white$\ell-\pi$  & 600   &0.39&0.66&7.2&20&700&0.68&3.2&1.5&23&140\\
\white$\ell-\rho$ &1200 & 1.4&1.7&130&74&340&2.5&64&3.8&190&1200\\
\hline \hline
\multicolumn{2}{|c||}{combined}\white&\bf 0.23&\bf 0.34&3.0&15&22&\bf 0.4
&1.4&\bf0.78&13&18\\
\hline
\end{tabular}

\bigskip
\noindent{\bf Table \tabtauf}{\rm\quad 1 s.d. accuracies with which
$\hat f_{1\tau}$ and
$\hat f_{2\tau}$ can be measured \\ for a given number $N$ of
events in various
$\tau$ decay modes}

\end{center}

\section{Conclusions}
In this article we have investigated the radiative decays
$\rm Z\to \ell^+\ell^-\gamma\ (\ell=\mu,\tau)$ with respect to
the possibility of performing tests of the CP symmetry. The
relevant CP-odd parameters are the EDM and WDM coupling parameters
of the leptons, $\tilde d^\gamma_\ell,\tilde d_\ell^{\rm Z}$ and
4-point coupling parameters $\hat f_{i\ell}$. For the muon case
the EDM does not contribute to CP-odd correlations in
Z decays to leading order. But
in the effective Lagrangian approach
 $\tilde d^\gamma_\mu$ can be
bounded by the limits obtained in the direct search for an
EDM of the muon. On the other hand, the radiative Z decays
offer the possibility to measure the muon's WDM and its
4-point couplings $\hat f_{1\mu},\hat f_{2\mu}$. The
accuracies obtainable with a realistic number of events are
collected in Tab. 3.

\medskip

For the $\tau$-lepton one can in principle measure all four CP-odd
parameters $\tilde d^\gamma_\tau,\tilde d_\tau^Z,
\hat f_{1\tau},\hat f_{2\tau}$ using radiative Z decays. The
measurement of $\hat d_\tau^{\rm Z}$ using about $10^4\ {\rm Z}\to\tau^+
\tau^-\gamma$ decays has a 1 s.d. sensitivity (cf. Tab. 4)

\begin{equation}
\delta\hat d_\tau^{\rm Z}=0.018.
\end{equation}

\noindent This is, as expected, weaker than the bound (\ref{13a}) obtained
from $\rm Z\to\tau^+\tau^-$ decays. Radiative Z decays offer
the possibility to perform a true measurement of the EDM coupling
parameter $\hat d^\gamma_\tau$. To date no direct determination
of $\hat d^\gamma_\tau$ exists. Another reaction where
$\hat d^\gamma_\tau$ can be measured  is $e^+e^-\to\tau^+\tau^-$
away from the  Z resonance (cf. \cite{BNO}).
With $10^6$ $\tau$-pairs at $\sqrt s=10$ GeV for instance
one gets a sensitivity:

\begin{equation}
\delta\hat d^\gamma_\tau = 0.043.
\end{equation}

\noindent 
This is to be compared to $\delta\hat d_\tau^\gamma=0.17$
or 2.2 in Tab. 4 using the optimal observable or $V^*$, respectively.
Finally, the parameters $\hat f_{1\tau}, \hat f_{2\tau}$ are
special to $Z\to\tau^+\tau^-\gamma$ decays and can be
measured with accuracies of order 0.5 with the number of events
given in Tab. 5.

\medskip

An investigation of radiative Z decays can thus give information
of CP-odd coupling parameters of leptons which is complementary
to the one obtainable from other sources. 
We have found that for the observables and cuts we considered
correlations between the CP--odd coupling parameters
are negligible. Thus we investigated each coupling parameter separately.
In an analysis of real experimental data other cuts may be used
and smearing due to detector effects will play a role.
One will again be confronted with the question of correlations
and in a sophisticated
analysis one will have to consider the parameters
$\hat f_{1\ell}, \hat f_{2\ell}, \hat d_\ell^{\gamma},
\hat d_\ell^{\rm Z}$ simultaneously
for each lepton $\ell$. It should then be advantageous to use 
couplings and observables which give diagonal covariance matrices.
The general methods to achieve this which are explained in 
\cite{DN} can easily be adapted for our case here.

\medskip

Independently of CP-violation
it should be interesting to study the reaction $\rm Z\to\ell^+\ell^-
\gamma$ in the phase space region where the photon is ``hard''. Any new
effects in lepton physics, like substructures or excited leptons,
could lead to contact interactions between $\rm Z, \ell$ and $\gamma$.
These should best show up in ``hard'' 3-body decays, i.e. in regions of
phase space where the photon has high energy and is well separated from
$\ell^+$ and $\ell^-$. There the SM contribution is relatively small.
An investigation of the ratio

\begin{equation} \label{R}
R(y) = 
{ \Gamma(\rm Z\to\ell^+\ell^-\gamma) \over \Gamma_{\rm SM}
(\rm Z\to\ell^+\ell^-\gamma) }
\end{equation}

\noindent
as function of the cut parameter $y$ (\ref{ycut})
should already be revealing. As an example we have
calculated this ratio for our ansatz SM plus ${\cal L}_{\rm CP}$
(\ref{LCP})
assuming that only $f_{\rm V\ell}$ and $f_{\rm A\ell}$ are different
from zero. To a very good approximation (cf.\cite{Haberl}) 
these couplings add to $\Gamma({\rm Z}\to\ell^+\ell^-)$ a term
$\Delta\Gamma_{\rm CP}(f_{\rm V\ell}, f_{\rm A\ell}; y)$ proportional
to $\hat f_{\rm V\ell}^2 + \hat f_{\rm A\ell}^2$.
Thus we plot 

\begin{equation} \label{Ry1}
{R(y)-1 \over \hat f_{\rm V\ell}^2 + \hat f_{\rm A\ell}^2 }
=
{\Delta\Gamma_{\rm CP}(\hat f_{\rm V\ell}, \hat f_{\rm A\ell}; y) \over
\Gamma_{\rm SM} (\rm Z\to \ell^+\ell^-\gamma)
( \hat f_{\rm V\ell}^2 + \hat f_{\rm A\ell}^2)
}
\end{equation}

\noindent in Fig. 4 (solid line). We see that a value
$\hat f_{\rm V\ell}^2 + \hat f_{\rm A\ell}^2 = 1$ will lead to
deviations of $R(y)$ from 1 of up to about
10\% at $y=0.2$. An investigation of 

\begin{equation} \label{r}
 r(y) = {
\Delta\Gamma_{\rm CP}(\hat f_{\rm V\ell}, \hat f_{\rm A\ell}; y) \over
\Delta\Gamma_{\rm CP}(\hat f_{\rm V\ell}, \hat f_{\rm A\ell}) },
\end{equation}

\noindent plotted in Fig. 4 (dashed line) shows that the {\it absolute}
contribution of CP--odd parameters to the width
decreases rapidly for increasing values of the cut parameter $y$. 
Here $\Delta\Gamma_{\rm CP} (\hat f_{\rm V\ell}, \hat f_{\rm A\ell})$ 
is the addition
to the width for cut parameter $y=0$ as given in (\ref{deltagammaf}).

\medskip

The calculations reported in this article required a considerable amount
of symbolic and numerical computation.
We used the new programming language {\it M} \cite{M} for the symbolic
manipulations including the expansion of traces of Dirac--matrices, 
simplification of the resulting large expressions, and  generation
of optimized Fortran program code. The numerical phase space 
integrations (with dimensions up to 11) were done with the Fortran routine
VEGAS. The programs which we used as well as Monte Carlo event 
generators for the reactions discussed in this article can be
obtained from the authors (World Wide Web address:
{\tt http://www.thphys.uni-heidelberg/\~\ \hskip-3pt overmann}).

\bigskip\bigskip

{\it Acknowledgements \quad} We are grateful to
W. Bernreuther, M. Diehl, 
P. Haberl,
N. Wermes,
and M. Wunsch
for many useful and informative discussions and to W. Bernreuther,
M. Diehl,  and
P. Haberl
for a careful reading of the manuscript.

\newpage

\noindent{\bf Figure Captions}

\vskip 2cm

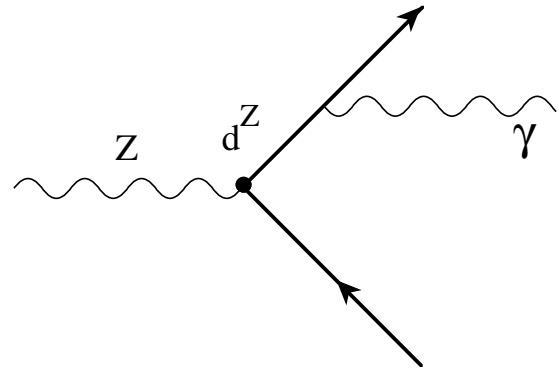
\noindent {\bf Figure 1}

\noindent
Feynman diagrams  contributing  to $\rm Z \to \ell^+\ell^-\gamma$:

a) One of the lowest order Standard Model diagrams

b) Contribution of $f_{\rm A\ell}$ and  $f_{\rm V\ell}$

c) Contribution of the electric dipole moment $\tilde d^\gamma_\ell$

d) Contribution of the weak dipole moment $\tilde d^{\rm Z}_\ell$

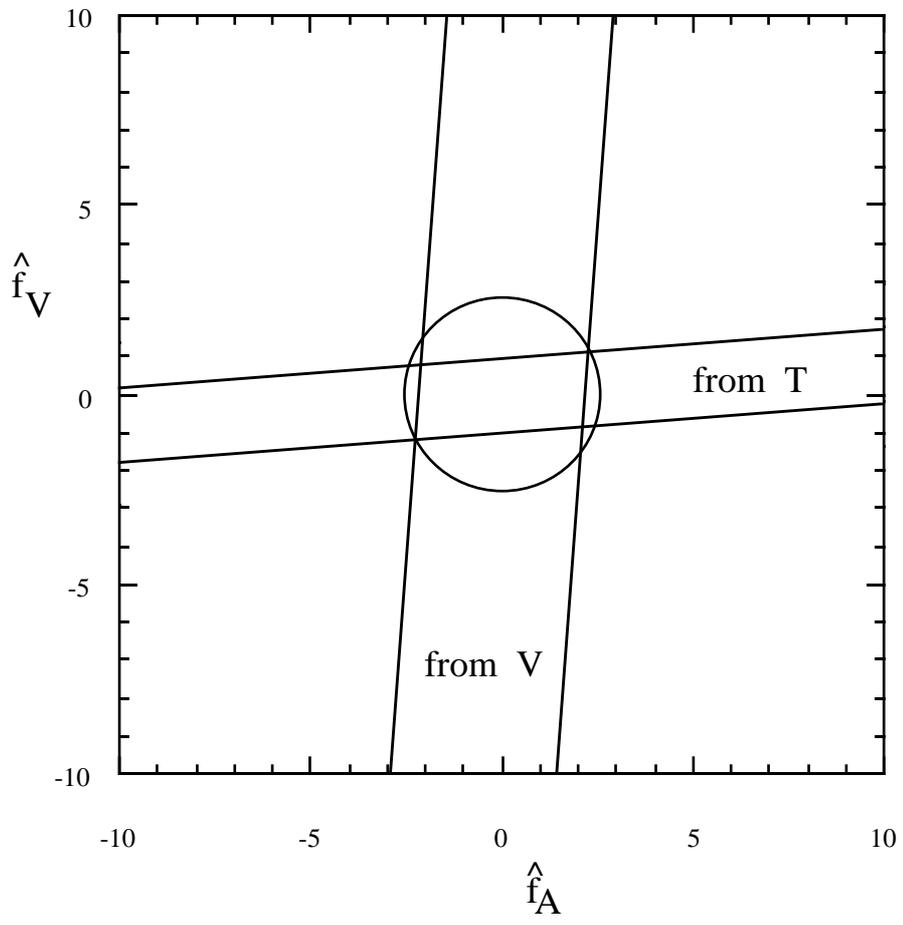
\bigskip\bigskip\noindent {\bf Figure 2}

\noindent Contour plot of the 2 s.d. errors on
 $\hat f_{\rm V\mu}, \hat f_{\rm A\mu}$ obtainable ideally
from the
measurement of the correlations $T,V$ from Tab. 3  (stripes) and from the contribution to the width (circle) (c.f. (\ref{10a},\ref{findirect}))
assuming 10000
$\mu^+\mu^-\gamma$ events and zero mean values.

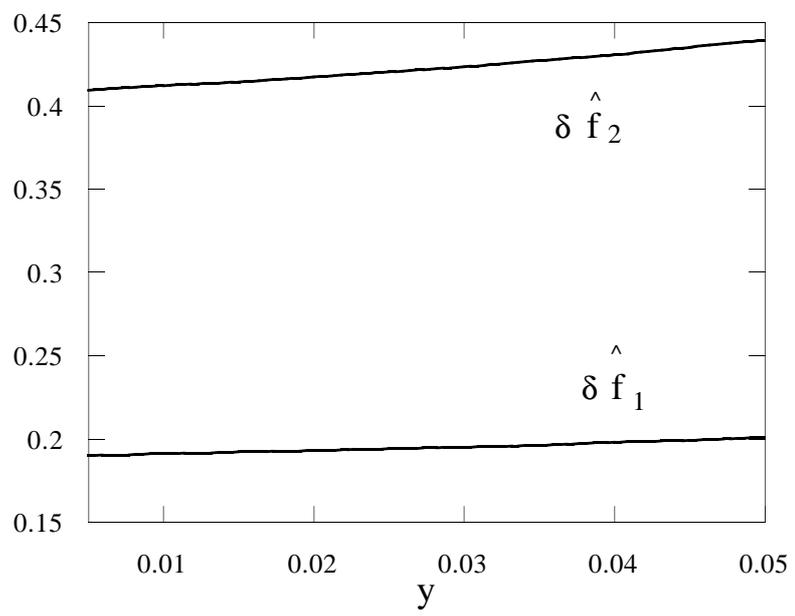
\bigskip\bigskip \noindent {\bf Figure 3}

\noindent
1 s.d. accuracies for the determination of $\hat f_{1\mu}$ and
$\hat f_{2\mu}$ for different values of the phase space cut parameter $y$.
Here we assume a number of  $\mu^+\mu^-\gamma$ events corresponding 
to $10^4$ for $y = 0.03$.

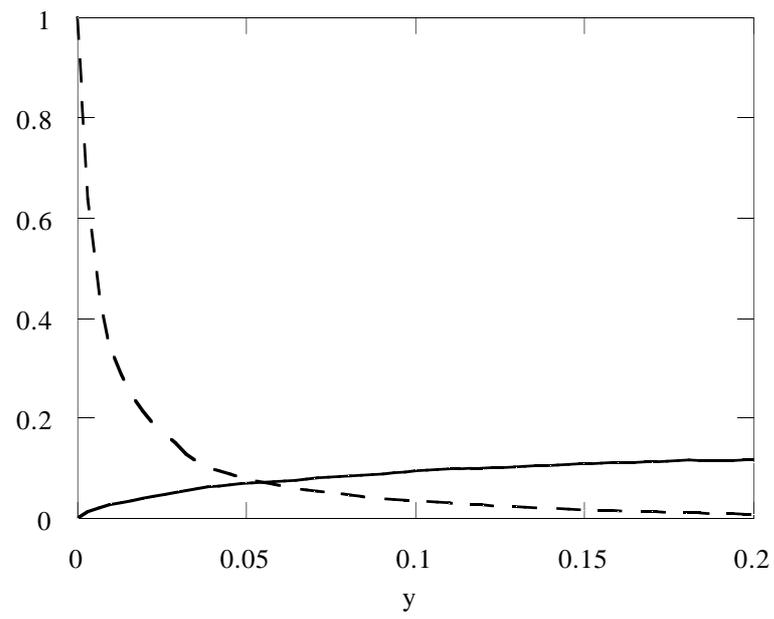
\bigskip\bigskip \noindent {\bf Figure 4}

\noindent The ratio $(R(y)-1)/(\hat f_{\rm V\ell}^2 + \hat f_{\rm A\ell}^2)$ 
 (solid line) as given in  (\ref{R},\ref{Ry1}), 
and the ratio $r(y)$ (dashed line)
defined in (\ref{r}) as function of the cut parameter
$y$ (\ref{ycut}).

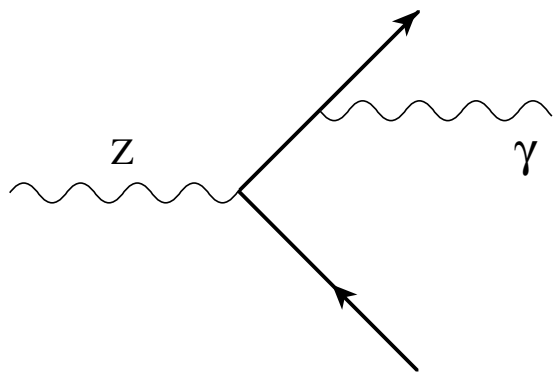
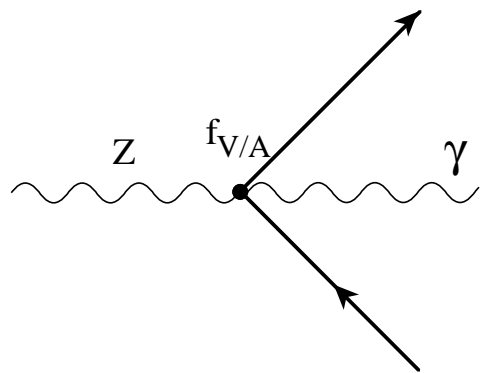
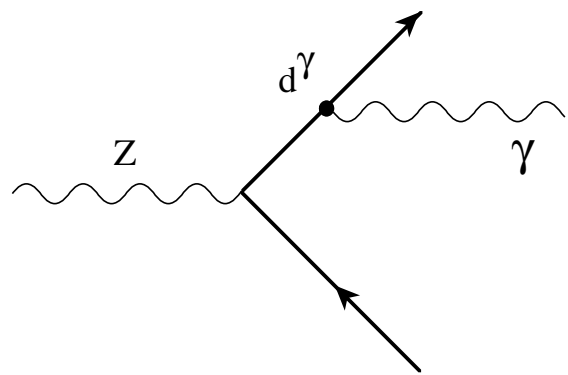
\end{document}